\documentclass[12pt]{iopart}
\usepackage{amssymb}
\usepackage{graphicx}
\usepackage{dcolumn}
\usepackage{bm}

\begin{document}
\title[Reconciling cyanobacterial metabolite and transport experiments with modelling]{Reconciling cyanobacterial fixed-nitrogen distributions and transport experiments with quantitative modelling} 

\author{Aidan I Brown and Andrew D Rutenberg}
\address{Department of Physics and Atmospheric Science, Dalhousie University, Halifax, Nova Scotia, Canada B3H 4R2}
\ead{\mailto{aidan@dal.ca}, \mailto{andrew.rutenberg@dal.ca}}

\date{\today}

\begin{abstract} 
Filamentous cyanobacteria growing in media with insufficient fixed nitrogen differentiate some cells into heterocysts, which fix nitrogen for the remaining vegetative cells. Transport studies have shown both periplasmic and cytoplasmic connections between cells that could transport fixed-nitrogen along the filament.  Two experiments have  imaged fixed-nitrogen distributions along filaments.  In 1974, Wolk \emph{et al} found a peaked concentration of fixed-nitrogen at heterocysts using autoradiographic techniques. In contrast, in 2007, Popa \emph{et al}  used nanoSIMS  to show large dips at the location of heterocysts, with a variable but approximately level distribution between them.  With  an integrated model of fixed-nitrogen transport and cell growth, we recover the results of both Wolk \emph{et al} and of Popa \emph{et al} using the same model parameters. To do this, we account for immobile incorporated fixed-nitrogen and for the differing durations of labeled nitrogen fixation that occurred in the two experiments.  The variations seen by Popa \emph{et al} are consistent with the effects of cell-by-cell variations of growth rates, and mask diffusive gradients.  We are unable to rule out a significant amount of periplasmic fN transport.
\end{abstract}


\maketitle
\section{Introduction}

Filamentous cyanobacteria are model  multicellular  organisms that form unbranched filaments of clonal cells.   In conditions of low exogenous fixed-nitrogen, terminally-differentiated heterocyst cells develop along the filament separated by clusters of photosynthetic vegetative cells \cite{flores10,kumar09}. Heterocysts fix dinitrogen \cite{wolk74} and provide the fixed-nitrogen (fN) to vegetative cells where it accommodates ongoing growth.  Vegetative cells cannot themselves fix nitrogen due to oxygen produced by photosynthesis \cite{meeks02}.  Because fN transport is required for ongoing growth, and because local fN deprivation is a key early signal for heterocyst development, it is important to understand how fN is transported within the cyanobacterial filament. Since fN is a metabolite many convenient genetic visualization techniques are not directly applicable.  Nevertheless, radio-isotopes of nitrogen were used to show that fN is rapidly incorporated into glutamine (molecular mass 146 Da) within heterocysts \cite{thomas77,wolk76}. Fixed-nitrogen is likely transported to vegetative cells as glutamine or other amino acids \cite{flores10,meeks02}.  Various imaging approaches have indicated cytoplasm-to-cytoplasm connections between adjacent cells and a shared contiguous periplasmic space, either of which might transport fN between cells \cite{haselkorn08}.  

Early indications for direct cytoplasmic connections between adjacent vegetative cells within cyanobacterial filaments were electron-microscopy images of small ( $\approx 50 \AA$ diameter) holes in the septal cell membrane called microplasmodesmata \cite{haselkorn08,giddings78,giddings81}.  Larger pores have also been observed connecting heterocysts to adjoining vegetative cells \cite{lang71}. Recently, direct cytoplasmic transport of the small exogenous fluorophore calcein (molecular mass 623 Da) was observed in \emph{Anabaena} sp. strain PCC7120 (hereafter PCC7120) using fluorescence recovery after photobleaching (FRAP) \cite{mullineaux08}. After photobleaching of a cell, fluorescence recovered on a timescale of tens of seconds ---  allowing transport constants to be estimated. Since calcein is not native to cyanobacteria, it is thought that direct cytoplasmic transport is nonspecific and could be used to transport fN.  The protein SepJ (FraG) appears to play a key role in direct cytoplasmic transport. SepJ includes a coiled-coil domain that prevents filament fragmentation and a permease domain required for diazotrophic growth \cite{flores11}. SepJ also appears to be needed for microplasmodesmata formation \cite{herrero07}.  There are no indications of ATPase activity of SepJ, or any other energy requirements for transport between cells --- indicating that direct cytoplasmic transport of small molecules such as fN between cells is likely to be by passive diffusion.  

In addition to direct cell-to-cell connections, electron microscopy images indicate that the periplasmic space of (Gram negative) cyanobacterial filaments is contiguous (see Flores \emph{et al} \cite{flores06}).   Supporting this, transport of periplasmic GFP (molecular mass $\approx 27 kDa$) has been recently observed in FRAP studies \cite{mariscal07}, though this is somewhat controversial \cite{kumar09,haselkorn08,zhang08}. Nevertheless, the outer membrane has been shown to provide a permeability barrier to fN compounds \cite{nicolaisen09}. Furthermore, knocking out amino-acid permeases (transporters) that act between the periplasm and cytoplasm leads to impaired diazotrophic growth \cite{montesinos95,picossi05,pernil08}.  Together these imply that there may be some amount of periplasmic fN present during diazotrophic growth, which would then passively diffuse between adjacent cells via the contiguous periplasmic space.  However, it is not clear how large diffusive fluxes are in the periplasm compared to those between cytoplasmic compartments. 

There have been two studies of nitrogen transport along cyanobacterial filaments \cite{wolk74,popa07}, but neither had the transverse spatial resolution to distinguish cytoplasm from periplasm. The first study was by Wolk \emph{et al} in 1974 \cite{wolk74} (hereafter simply ``Wolk''). Diazotrophically growing filaments of \emph{Anabaena cylindrica} were provided with dinitrogen gas composed of $^{13}$N (half-life of less than 10 minutes) for 2 minutes before being imaged. Radioactive decay tracks from $^{13}$N were used to approximate the distribution of labeled fN in the filament. Wolk found that the fN was peaked at the location of heterocysts --- consistent with diffusive transport away from a heterocystous source. 

A more recent study by Popa \emph{et al} in 2007 \cite{popa07} (hereafter simply ``Popa'') studied diazotrophically growing \emph{Anabaena oscillarioides} filaments with a nanoSIMS (nanometer-scale secondary ion mass spectrometry)  technique after they were provided with stable dinitrogen gas composed of  $^{15}$N. After four hours, they found \emph{dips} of the fN distribution at the locations of heterocysts, and a noisy plateau between them (see red curve in Fig.~\ref{fig:popa}c, below) --- with no evidence of a concentration gradient consistent with diffusive transport away from heterocysts. 

It has not been immediately clear how to reconcile the Wolk and Popa fN distributions. It is tempting, but not necessary, to attribute qualitative differences between the Wolk and Popa distributions to distinct species-specific physiology. Instead, we hypothesize that fN transport is qualitatively similar in all heterocystous filamentous cyanobacterial species and investigate the quantitative fN patterns expected from modelling the two experimental approaches with a single model.

Wolk \emph{et al} \cite{wolk74} developed a quantitative model of fN transport and incorporation along the cyanobacterial filament, including diffusive transport and fN consumption in vegetative cells. However, the Wolk model did not distinguish periplasmic from cytoplasmic transport, used a homogeneous continuum model of the filaments, did not include stochasticity, and assumed that cytoplasmic fN was always rate-limiting for growth --- i.e. growth vs local fN concentration was strictly linear.  We found (see Appendix A) that with this last assumption the Wolk model was unable to reproduce the distribution of Popa after prolonged production of fN at heterocysts.    More recently, a stochastic cellular computational model of cyanobacterial growth and fN transport was developed by Allard \emph{et al} \cite{allard07}. However, they assumed that only periplasmic fN transport was significant and they did not investigate the possibility of local fN starvation. We adapt their model, with both cytoplasmic and periplasmic transport but stripped of heterocyst development and supplemented by vegetative growth that is independent of non-zero fN concentrations, to the study of fN distributions. Using this model, we reconcile the two qualitatively distinct fN distributions of Wolk and of Popa. We then use the qualitative Popa results to constrain possible periplasmic fN transport. We find periplasmic fN transport is not necessary to explain the current data but cannot be ruled out as a significant contribution.

\section{Model}
\label{sec:model}

\subsection{Fixed-nitrogen transport}
\label{subsec:dynamics}

\begin{figure}
  \begin{center}
    \includegraphics[width=3.0in]{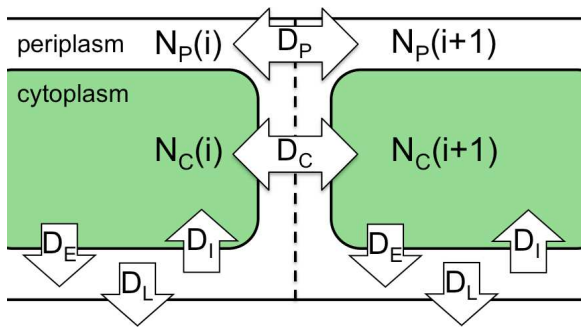} \\
  \end{center}
  \caption{\label{fig:nitrogendynamics} Schematic of fixed-nitrogen dynamics as represented by Eqns.~\protect\ref{eq:cytonitrogen} and \protect\ref{eq:perinitrogen}. Fixed-nitrogen amounts are indexed to each cell by $i$, with $N_C(i)$ the cytoplasmic fixed-nitrogen and $N_P(i)$ the periplasmic fixed-nitrogen of each cell. Each shaded green region represents the cytoplasms of two adjacent cells. $D_C$ describes the cytoplasm-cytoplasm transport, $D_P$ the periplasm-periplasm transport, $D_I$ the import from the periplasm to the cytoplasm, $D_E$ the export from the cytoplasm to the periplasm and $D_L$ the loss from the periplasm to outside of the filament.}
\end{figure}

Our fixed-nitrogen (fN) transport and incorporation model is similar to that of Allard \emph{et al} \cite{allard07}, and tracks the total amount of cytoplasmic fN, $N_C(i,t)$, in each cell $i$ vs. time $t$: 
\begin{equation}
	\label{eq:cytonitrogen}
	\frac{d}{dt}N_C(i,t)=\Phi_C(i)+D_I N_P(i,t)-D_E N_C(i,t)+G_i,
\end{equation}
where $\Phi_C(i) \equiv \Phi_{RC}(i-1)+\Phi_{LC}(i+1)-\Phi_{LC}(i)-\Phi_{RC}(i)$ is the net diffusive cytoplasmic flux into cell $i$ and is equal to the sum of the fluxes into the cell from the left and right minus the fluxes out of the cell to the left and the right.  Similarly we track the corresponding periplasmic fixed-nitrogen $N_P(i,t)$: 
\begin{equation}
	\label{eq:perinitrogen}
	\frac{d}{dt}N_P(i,t)=\Phi_P(i)-(D_I+D_L) N_P(i,t)+D_E N_C(i,t),
\end{equation}
where, similarly, $\Phi_P(i) \equiv \Phi_{RP}(i-1)+\Phi_{LP}(i+1)-\Phi_{LP}(i)-\Phi_{RP}(i)$ is the net diffusive periplasmic flux into the periplasm of cell $i$.  In addition to fluxes along the filament, $D_I$ is the coefficient for  periplasm-to-cytoplasm transport, $D_E$ is the coefficient for cytoplasm-to-periplasm transport, and $D_L$ is the coefficient for losses from the periplasm to the external medium. These transport processes are shown  schematically in Fig.~\ref{fig:nitrogendynamics}. Note that any fN transferred between cytoplasm and periplasm, with $D_I$ or $D_E$, shows up with opposite sign in Eqns.~\ref{eq:cytonitrogen} and \ref{eq:perinitrogen} --- so that the total amount of free fN is unchanged by the exchange.  In addition to these transport terms, there is a source/sink term $G_i$ in Eqn.~\ref{eq:cytonitrogen} that describes fN production and consumption in the cytoplasms of heterocysts and vegetative cells, respectively. This $G$ term is discussed more extensively below in Sec.~\ref{sec:growth}.

Diffusive transport between cells is described by the flux terms $\Phi$.  Following Allard \emph{et al} \cite{allard07} and Mullineaux \emph{et al} \cite{mullineaux08} we assume that they are limited by the junctions between cells. Supporting this, there is no evidence in the images of fN by Popa \emph{et al} \cite{popa07} or of calcein by Mullineaux \emph{et al} \cite{mullineaux08} of subcellular gradients. Fick's law states that net diffusive fluxes between two cells will be proportional to the density difference between the cells. Each cell will have two outgoing fluxes, one to the left neighbour $\Phi_L$ and one to the right neighbour $\Phi_R$, each proportional to the local density $N(i,t)/L(i,t)$ and to transport coefficients $D_C$ or $D_P$ for cytoplasmic or periplasmic fluxes, respectively. Each cell will also have two incoming fluxes from its neighbouring cells.  

The value of the transport coefficients $D_C$ or $D_P$ will be proportional to the microscopic diffusivity of fN, but will also depend on the size and shape of cytoplasmic or periplasmic connections between cells --- which are not well characterized.  Nevertheless, we can take the cytoplasm-to-cytoplasm transport coefficients for calcein \cite{mullineaux08} and simply scale them to glutamine under the assumption that contributions due to connection geometry are unchanged.  Calcein transport coefficients reported by Mullineaux \emph{et al} \cite{mullineaux08} assumed a constant cell length. Multiplying their $E$ in units of $s^{-1}$ by a  cell length of 3.38 $\mu$m, gives our transport coefficients in units of $\mu$m s$^{-1}$.  We then scale from calcein to glutamine with a factor of the cube-root of the ratio of molecular masses, assuming Stokes-Einstein diffusivity that inversely depends upon radius. We obtain $D_C$=1.54 $\mu$m s$^{-1}$ between two vegetative cells and $D_C$=0.19 $\mu$m s$^{-1}$ between a vegetative cell and a heterocyst --- we keep these values fixed in this paper.  Note that with uniform transport coefficients, such as among clusters of vegetative cells, the flux terms in in Eqn.~\ref{eq:cytonitrogen} or \ref{eq:perinitrogen} look like a discrete Laplacian, $ \nabla^2_d (N/L) \equiv N(i+1,t)/L(i+1,t)+N(i-1,t)/L(i-1,t)-2N(i,t)/L(i,t))$ and lead to the more familiar diffusion equation when coarse-grained (see Eqn.~\ref{eq:diff} and \ref{app:growth}, below).

We can estimate a lower bound of periplasmic fN transport with the GFP-FRAP data of Mariscal \emph{et al} \cite{mariscal07}. They observed a fluorescence recovery after photobleaching timescale of approximately $150$s \cite{mariscal07}. This corresponds to $D_{Pgfp} \approx L/150s \approx  0.02 \mu$m s$^{-1}$, using a cell length $L \approx 3 \mu$m.   Scaling the transport of GFP (27 kDa) to glutamine (146 Da) by the cube-root of the molecular masses \cite{nenninger10}, i.e. by a factor of (27 kDa/146 Da)$^{1/3} \approx 5.7$, and taking this as a lower bound, then gives $D_P \gtrsim 0.1$ $\mu$m/s.  However, the relatively large GFP may diffuse particularly slowly through periplasmic connections compared to glutamine. FRAP experiments with GFP in \emph{Anabaena} PCC 7120 showed recovery in part of a single cell's periplasm after around 60s \cite{zhang08} and after around 5s \cite{mariscal07}. This faster time of 5s is similar to the 4s recovery time observed in the periplasm of \emph{E. coli} \cite{mullineaux06}, which also measured an associated diffusivity of $D_{coli}\approx 2.6 \mu$m$^2/s$. We use this diffusivity to determine an upper bound of $D_P$ by scaling it to glutamine and dividing by cell length to obtain $D_P = D_{coli} \times 5.7 / 3.38 \mu$m $ \approx 4.4 \mu$m/s.  The broad range we consider for periplasmic transport is then $0.1$ $\mu$m s$^{-1}$ $\lesssim $ $D_P$ $\lesssim$ 10 $\mu$m s$^{-1}$.

When we consider periplasmic transport, we must also consider exchange between the cytoplasm and the periplasm --- as well as loss from the periplasm. We expect that uptake of fN from the periplasmic to the cytoplasmic compartment, controlled by $D_I$ in Eqns.~\ref{eq:cytonitrogen}  and \ref{eq:perinitrogen}, is an active process, controlled by e.g. ABC transporters \cite{pernil08}.  Nevertheless, we expect that the rate of import is proportional to the periplasmic density, $N_P/L$, and to the number of transporters, which will themselves be proportional to the cell length $\sim L$. Similarly, leakage of fN from the cytoplasmic to the periplasmic compartment,  controlled by $D_E$, will be proportional to the cytoplasmic density $N_C/L$ but also to  the amount of membrane or of (leaky) ABC transporters ($\sim L$) \cite{rees09}. The result is transport terms $D_I$ and $D_E$ that are independent of length $L$ and have units of s$^{-1}$. Using the transport data of Pernil \emph{et al} \cite{pernil08} for uptake of extracellular glutamine, we estimate (see Appendix B) $D_I$ = 4.98 s$^{-1}$ and $D_E$ = 0.498 s$^{-1}$. We also consider the possibility of stronger transport, with $D_I$ = 49.8 s$^{-1}$ with $D_E$ = 4.98 s$^{-1}$, because the outer membrane permeability barrier \cite{nicolaisen09} may decrease extracellular uptake with respect to the periplasmic uptake $D_I$.  There is evidence that fixed-nitrogen is lost from the filament to the external medium \cite{paerl78,thiel90} and we include a loss term for the periplasm with coefficient $D_L$ in the Eq.~\ref{eq:perinitrogen}. We take the loss rate as very small compared to the reuptake rate, using 1$\%$ of the lower $D_I$ value i.e.  $D_L$ = 0.0498 s$^{-1}$  --- qualitatively corresponding to the observation of a permeability barrier in the outer membrane \cite{nicolaisen09}. Our results are qualitatively unchanged in the range of 0 $ \leq D_L/D_I \lesssim $ 0.1.

\subsection{Cell Growth and Division}
\label{sec:growth}
We take PCC7120 cells to have a minimum size of $L_{min}$ = 2.25$\mu$m and a maximum size of $L_{max}=2 L_{min}$ \cite{flores10,kumar09}.  When a cell reaches $L_{max}$ it is divided into two cells of equal length, each of which is randomly assigned a new growth rate and half of the fixed nitrogen in the parent cell. We found that average concentration profiles were largely insensitive to stochastic effects. Accordingly, most of our results came from a deterministic model where all cells were initialized with the same length $L_0=2.8$$\mu$m and the same doubling time $T_D = 20$h \cite{picossi05,herrero90}. For our stochastic results, presented in Figs.~\ref{fig:popa}(b) and (c), we initialized lengths randomly from an analytical steady state distribution of cell lengths  ranging between $L_{min}$ and $L_{max}$ \cite{powell56}. We define a minimum doubling time $T_{min}$ = $T_D-\Delta$ and a maximum doubling time $T_{max}$ = $T_D+\Delta$. These then define a minimum growth rate $R_{min}= L_{min}/T_{max}$ and a maximum growth rate $R_{max} =  L_{min} / T_{min}$ and we randomly and uniformly select optimal growth rates $R_i^{opt}$ from the range $R_i\in\left[R_{min},R_{max}\right]$. The standard deviation of the maximal growth rate is then $\sigma_R = \sqrt{2/3} \Delta L_{min} /[(T_D+\Delta)(T_D-\Delta)]$. We expect that the coefficient of variation of the growth rate $\sigma_R/R_{avg}$ is of similar magnitude to that seen in a study of mutant \emph{Anabaena} \cite{allard07}, which was $\sigma_R/R_{avg} \approx 0.165$. 

In Eqn.~\ref{eq:cytonitrogen},  for vegetative cells $G_i=G_{veg}$ is a growth term determining the rate of freely diffusing cytoplasmic fN removed to support growth of the cell.  $G_{veg}$ depends on the actual growth rate $R$ of the cell (in $\mu m/s$), which in turn depends upon the locally available cytoplasmic fN: 
\begin{equation}
	\label{eq:gequation}
	G_{veg}=-g R(R_i^{opt},N_C(i,t)),
\end{equation}
where $g$ is the amount of fixed nitrogen a cell needs to grow per $\mu$m -- which we now estimate. Dunn and Wolk \cite{dunn70} measure a dry mass for one cell of \emph{A. cylindrica} of 1.65$\times$10$^{-11}$g. Cobb \emph{et al} \cite{cobb64} measured the nitrogen content of \emph{A. cylindrica} to be 5-10$\%$ of dry mass (see also \cite{flores94}). We take their typical value of 10$\%$  \cite{cobb64} and scale it by volume from \emph{A. cylindrica} to PCC7120. \emph{A. cylindrica} appears to be 1.5$\times$ as wide and 1.5$\times$ as long as PCC7120 \cite{flores10}, giving $\approx$ 2.07$\times$10$^{10}$ N atoms per average cell of PCC7120. According to Powell's steady state length distribution of growing bacteria \cite{powell56} the average cell size is 1.44$\times$ as long as the smallest cell, so that $\approx$ 1.4$\times$10$^{10}$ N atoms are needed for a newly born cell to double in length. This implies that $g = 1.4 \times 10^{10}/L_{min} \simeq  6.2 \times10^9 \mu$m$^{-1}$.

We assume that cells increase their length at their optimal growth rate $R_i^{opt}$ as long as there is available cytoplasmic fixed-nitrogen (see discussion in Sec. \ref{sec:summary}). Otherwise, they can only grow using the fixed-nitrogen flux into the cell from neighbouring cells and the periplasm:
\begin{equation}
\label{eq:requation}
R=
   \cases{
      R_i^{opt}, & if $N_C(i,t)>0$\cr
      min(\Phi_{in}/g, R_i^{opt}) & if $N_C(i,t)=0$\cr
   }
\end{equation}
Note that cells with $N_C=0$ may still grow, but will be limited by the incoming fluxes of fixed-nitrogen from adjoining cells and the periplasm, $G_{veg}=\Phi_{in} = \Phi_{RC}(i-1) + \Phi_{LC}(i+1) + D_I N_P(i)$. When the incoming flux exceeds the requirement of maximal growth, then the cytoplasmic nitrogen concentration will rapidly become non-zero since we always have $R \leq R_i^{opt}$. 

In Eqn.~\ref{eq:cytonitrogen}, for heterocyst cells $G_i=G_{het}$.   The heterocyst fN production rate $G_{het}$ is chosen to supply the growth of approximately 20 vegetative cells and is  3.15$\times$10$^6$ $s^{-1}$ unless otherwise stated.

\subsection{Some numerical details}
\label{subsec:details}

Filaments were initiated with two heterocysts, separated on either side by vegetative cells, on a periodic loop.  Periodic boundary conditions were used to minimize end effects. Vegetative cells were then allowed to grow and divide, subject to available fixed-nitrogen.  Because of strong spatial gradients of free (unincorporated) fixed-nitrogen between heterocysts, we group and report our data with respect to the number of vegetative cells between two heterocysts --- typically we took a separation of 20 cells. This is approximately twice the typical heterocyst separation \cite{flores10,kumar09}, which allows us to investigate fixed-nitrogen depletion at the midpoint --- corresponding to the location of an intercalating heterocyst. 

We distinguish free from incorporated fixed-nitrogen (fN). Free fN is freely diffusing in the cytoplasm, and is simply given by $N_C(i,t)$.  We report this as a linear density $\rho_F \equiv N_C(i)/L_i$. Incorporated fN is the fixed-nitrogen that has been incorporated by cellular growth through the growth term $G_{veg}$.  We report the incorporated concentration $\rho_I \equiv \int G_{veg} dt/L_i$, where the growth is integrated over a fixed duration.  During cell division, half of any previously incorporated fN is assigned to either daughter cell. Note that since we are interested in isotopically labeled fixed-nitrogen \cite{wolk74,popa07}, we set all $N_C(i)=0$ at the start of our data gathering ($t=0$), corresponding to the introduction of isotopically labeled dinitrogen to the filament that is then fixed by the heterocysts and subsequently supplied to the filament as free fixed-nitrogen via $G_{het}$. The total fN, for e.g. Fig.~\ref{fig:fluxes}, in a cell is $\rho_T \equiv \rho_F+N_P/L + \rho_I$, where we include periplasmic contributions as well.

\section{Results and Discussion}
\label{sec:results}

\begin{figure*}
 \begin{center}
  \begin{tabular}{cc}
    \includegraphics[width=3.0in]{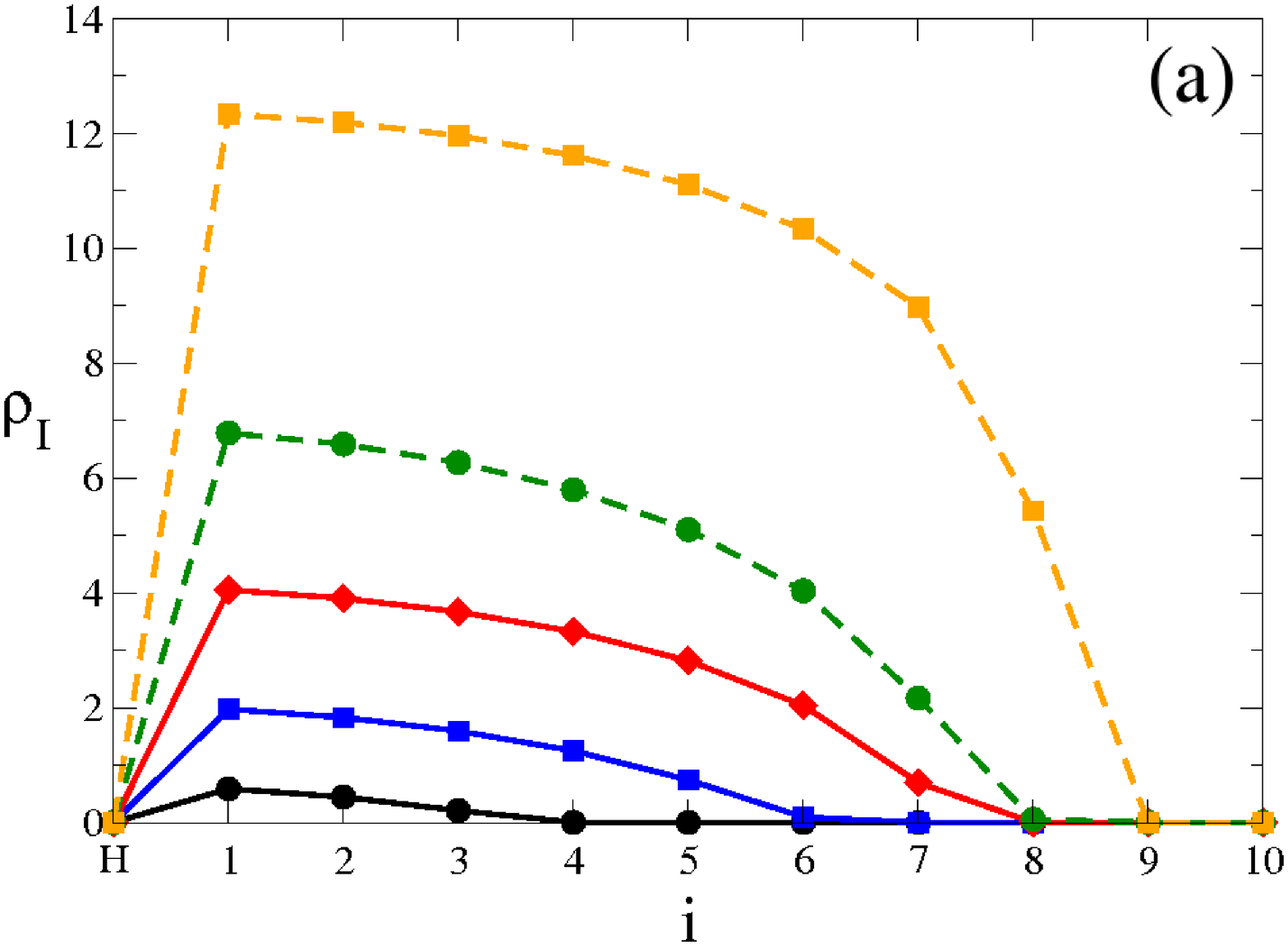}&
    \includegraphics[width=3.0in]{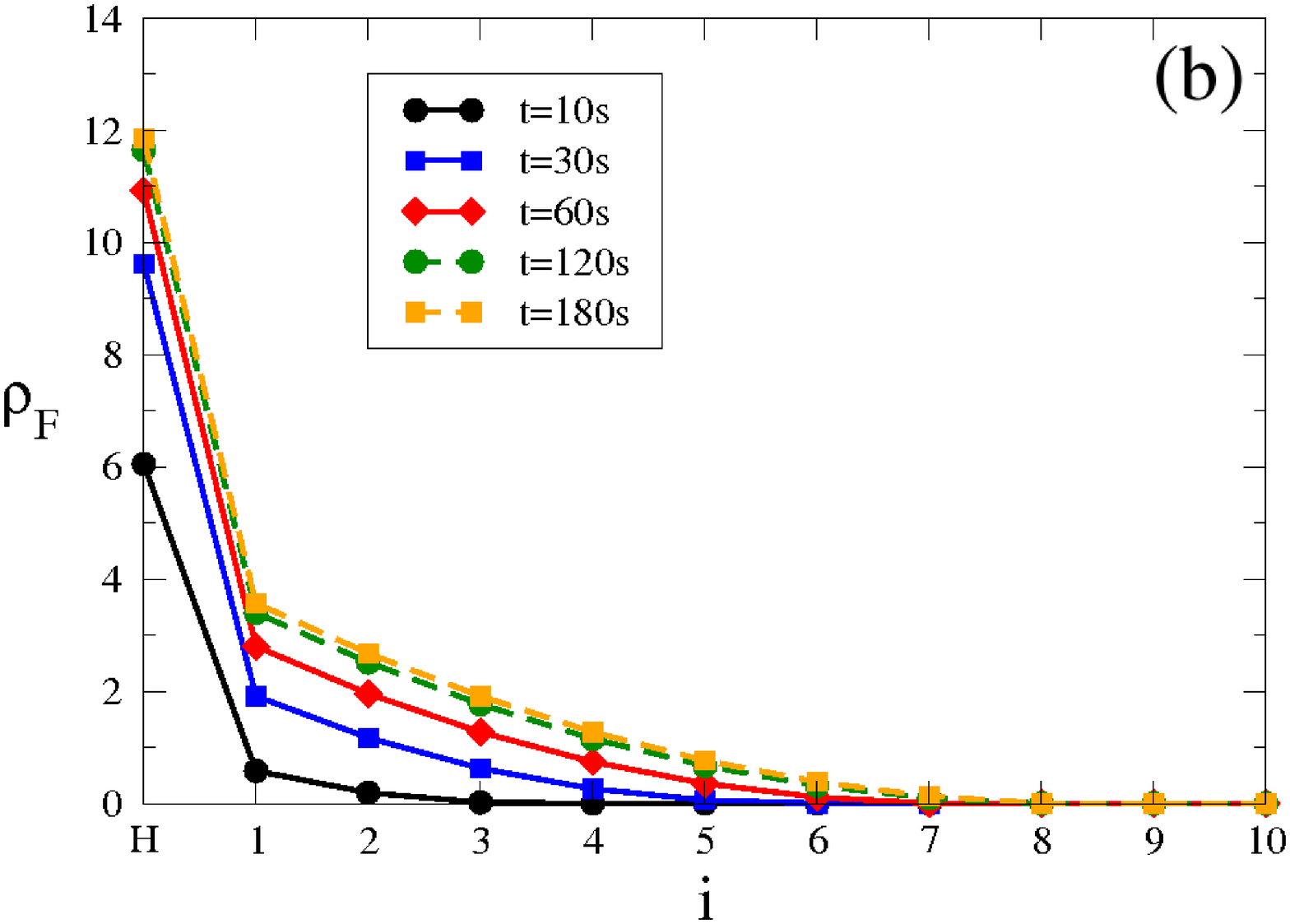}
  \end{tabular}
  \end{center}
  \begin{center}
  \begin{tabular}{cc}
    \includegraphics[width=3.0in]{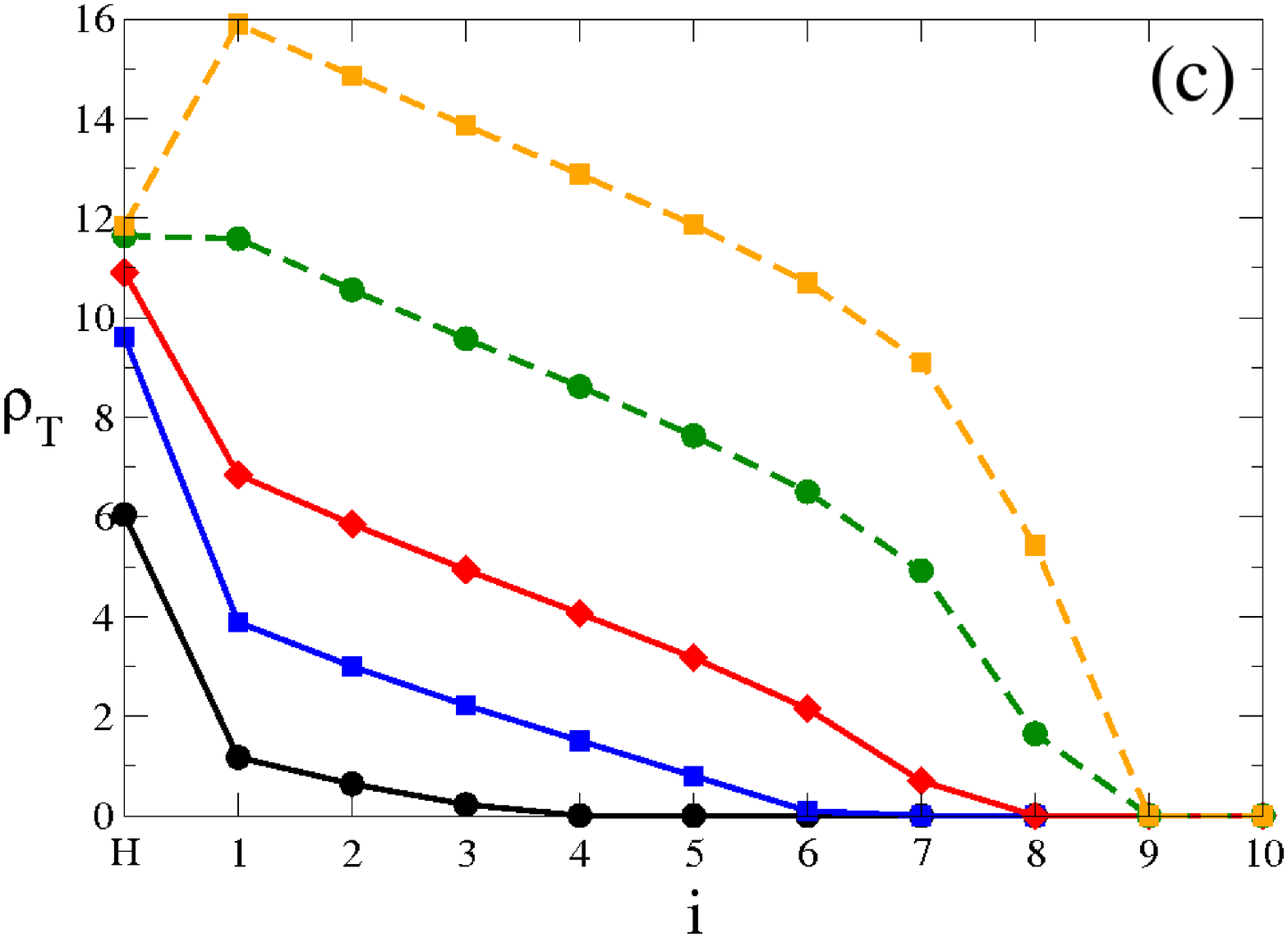}&
    \includegraphics[width=3.25in]{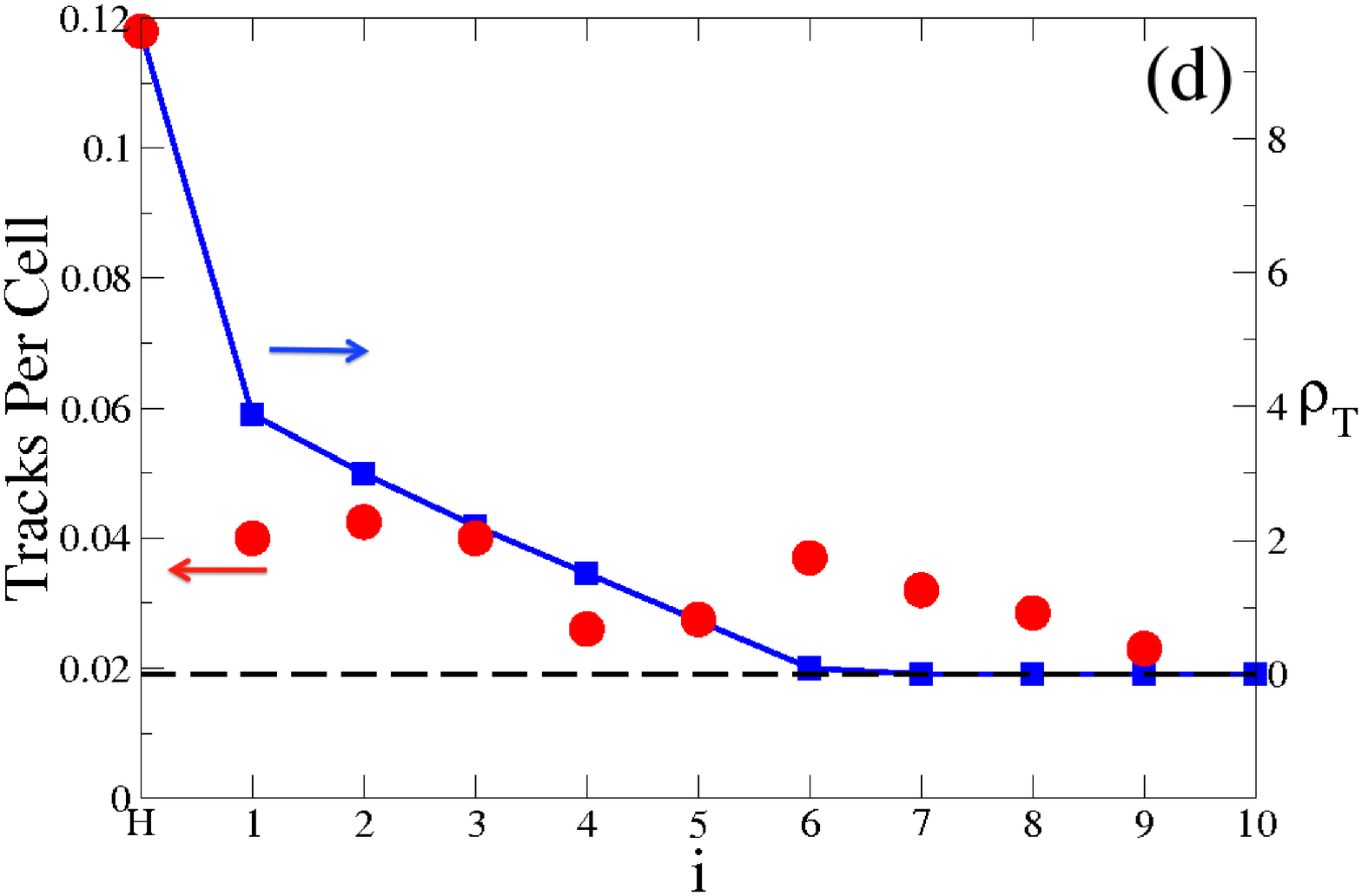}
  \end{tabular} 
  \end{center}
  \caption{\label{fig:wolk} Model fixed-nitrogen (fN) concentration vs. cell index $i$ (the number of cells from a heterocyst at $i=0$), for short intervals of labeled nitrogen fixation ranging from $10s$ to $180s$ as indicated by the legend in (b). The incorporated fN $\rho_I$ is shown in (a), the free fN $\rho_F$ in (b), and the total fN $\rho_T$ in (c). In (d) the red circles indicate the experimental data digitized from Wolk \emph{et al} \cite{wolk74}, and the blue squares are the $t=30s$ $\rho_T$ data from (c) (using the right axis, scaled and shifted to best agree with experimental data).  Our model data is from our deterministic model with no periplasmic transport, i.e. $D_C$ = 1.54 $\mu$m s$^{-1}$ and $D_I$ = $D_E$ = $D_P$ = 0.}
\end{figure*}

\subsection{Short-time fixed-nitrogen distributions}
\label{subsec:shorttime}

We first examined systems with only cytoplasmic transport (i.e. $D_I= D_E=D_P=0$) at short times comparable to those of Wolk \emph{et al} \cite{wolk74}. In Fig.~\ref{fig:wolk} we show fixed-nitrogen (fN) distributions along the filament, with respect to distance in cells from the nearest heterocyst, $i$. The cytoplasmic transport parameter was held constant at  $D_C$ = 1.54 $\mu$m/s,  estimated from calcein transport values in Section \ref{subsec:dynamics}. We  varied the time-interval between the start of labeled nitrogen fixation ($t=0$) and the times ($t=10s$, $30s$, $60s$, $120s$, and $180s$ --- see legend in panel $(b)$) at which the fN concentration profile along the filament was considered. As shown in (a), the incorporated fN, $\rho_I$, increases steadily over time for all vegetative cells reached by free fN, while incorporated fN remains at zero in the non-growing heterocyst (``H'').  As shown in (b), the free fN, $\rho_F$, spikes at the heterocyst, where it is produced, and progressively decreases with distance away from the heterocyst.  At later times, the free fN approaches a steady-state --- the distribution does not change significantly with time.  We show in (c) the total fN, $\rho_T= \rho_I+\rho_F$. At very short times the free fN dominates the total, resulting in the spike in the distribution at the heterocyst. At longer times the incorporated fN dominates the total --- which then loses its spike at the heterocyst and instead begins to form a dip at the heterocyst (see below).  
 
The radiographic technique of Wolk \emph{et al} \cite{wolk74} does not distinguish between free and incorporated fN, so it is appropriate to compare it with the total fN $\rho_T$ from Fig.~\ref{fig:wolk}(c). In (d) we show the average of the two distributions determined by Wolk \emph{et al} \cite{wolk74}, with filled red circles (left scale). We superimpose a qualitatively similar $\rho_T$ curve at $t=30s$ with the blue curve (right scale). We have shifted the $\rho_T=0$ axis to agree with the experimental background far from the heterocyst, and have adjusted the scale so that the curves agree at the heterocyst. The agreement is not bad, and captures the qualitative feature of the Wolk data --- namely the distinctive spike at the heterocyst.  This spike is a feature of short exposure to labeled fN. Indeed, while Wolk \emph{et al} took data at $t=120s$ we find our best qualitative agreement at $t=30s$.  In our model heterocysts immediately provide fN able to be transported. This is not the case in real heterocysts: glutamine levels in \emph{Anabaena cylindrica} linearly increase during the first minute of exposure to N$_2$, and other fixed nitrogen products appear after two minutes \cite{wolk76}. This one to two minute lag between exposure to labeled fN and its transport and incorporation is consistent with the 90s difference between the illustrative curves in Fig.~\ref{fig:wolk}(d).

\subsection{Steady-state fixed-nitrogen distributions}
\label{subsec:longtime}

\begin{figure*}
  \begin{center}
  \begin{tabular}{cc}
    \includegraphics[width=3.0in]{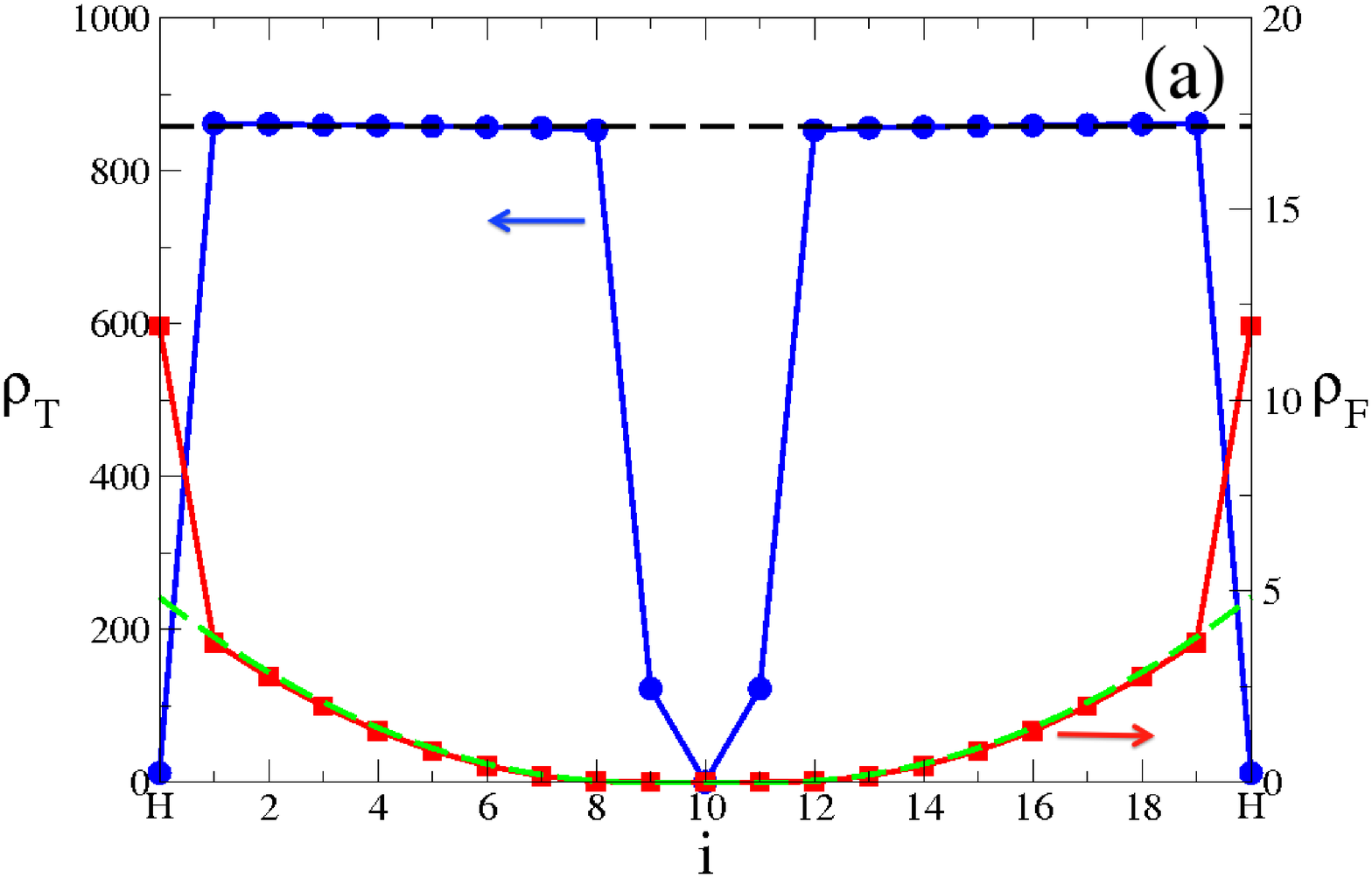}&
    \includegraphics[width=3.0in]{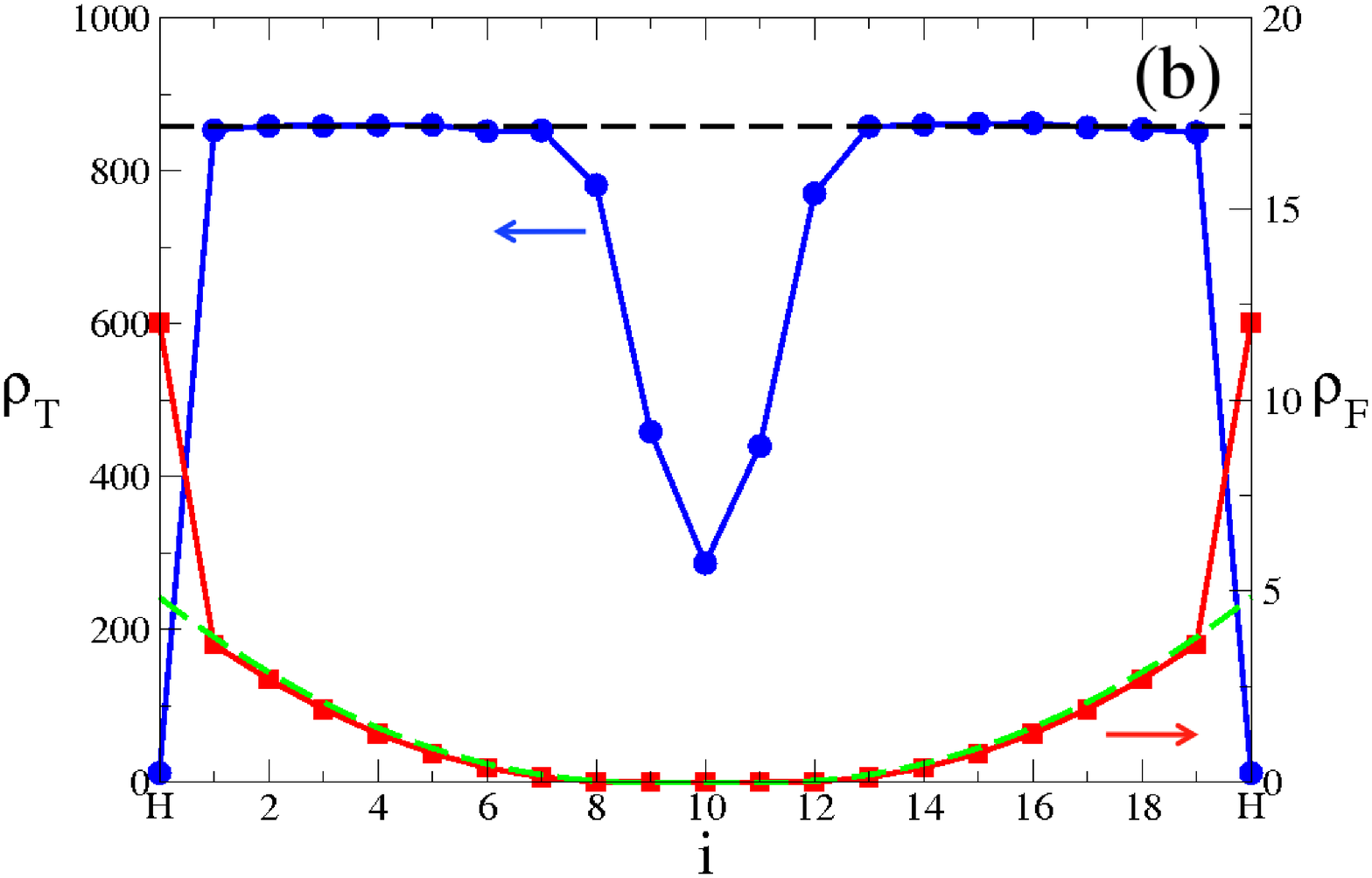}
  \end{tabular}
  \end{center}
  \begin{center}
  \begin{tabular}{cc}
    \includegraphics[width=3.0in]{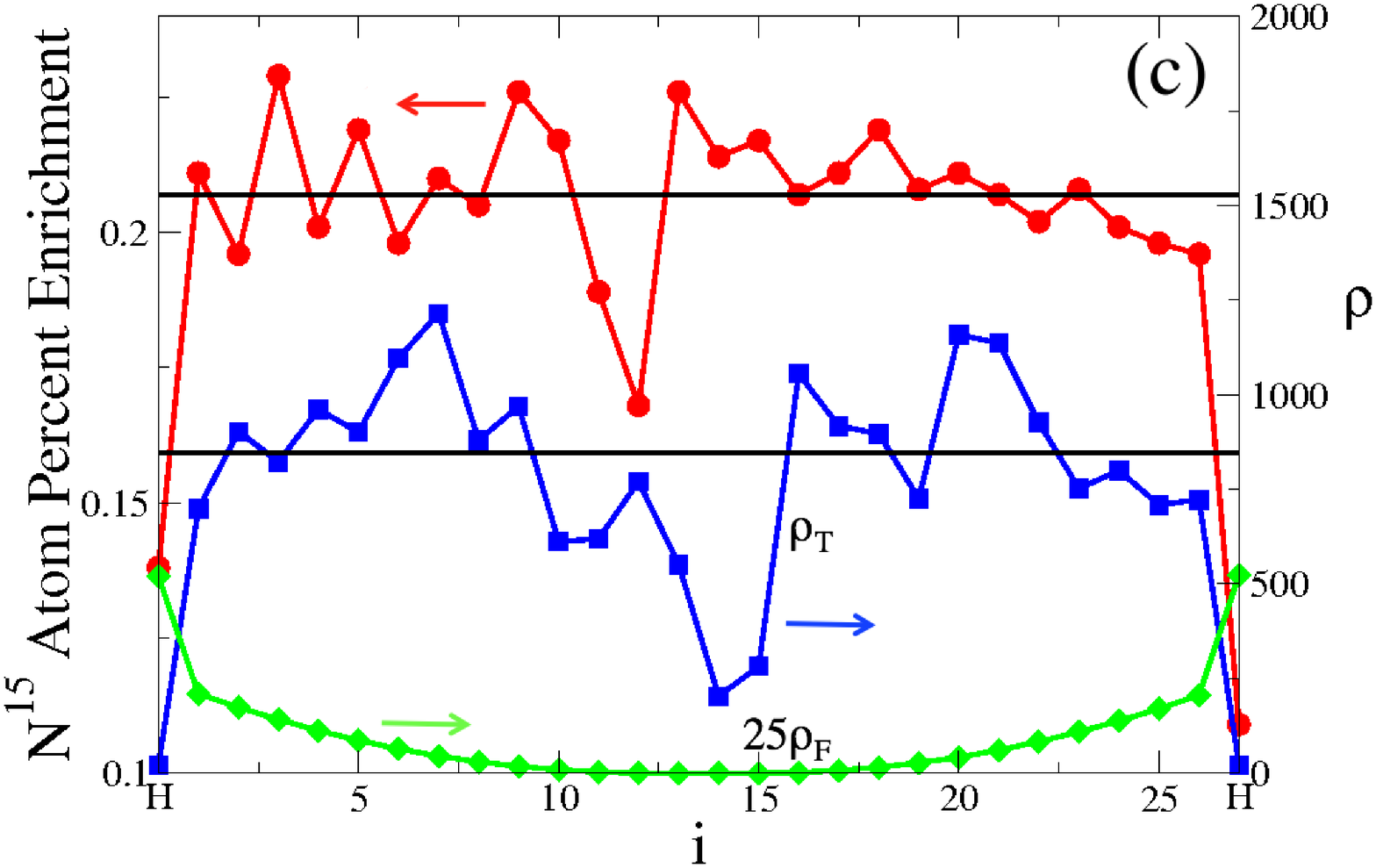}&
    \includegraphics[width=2.75in]{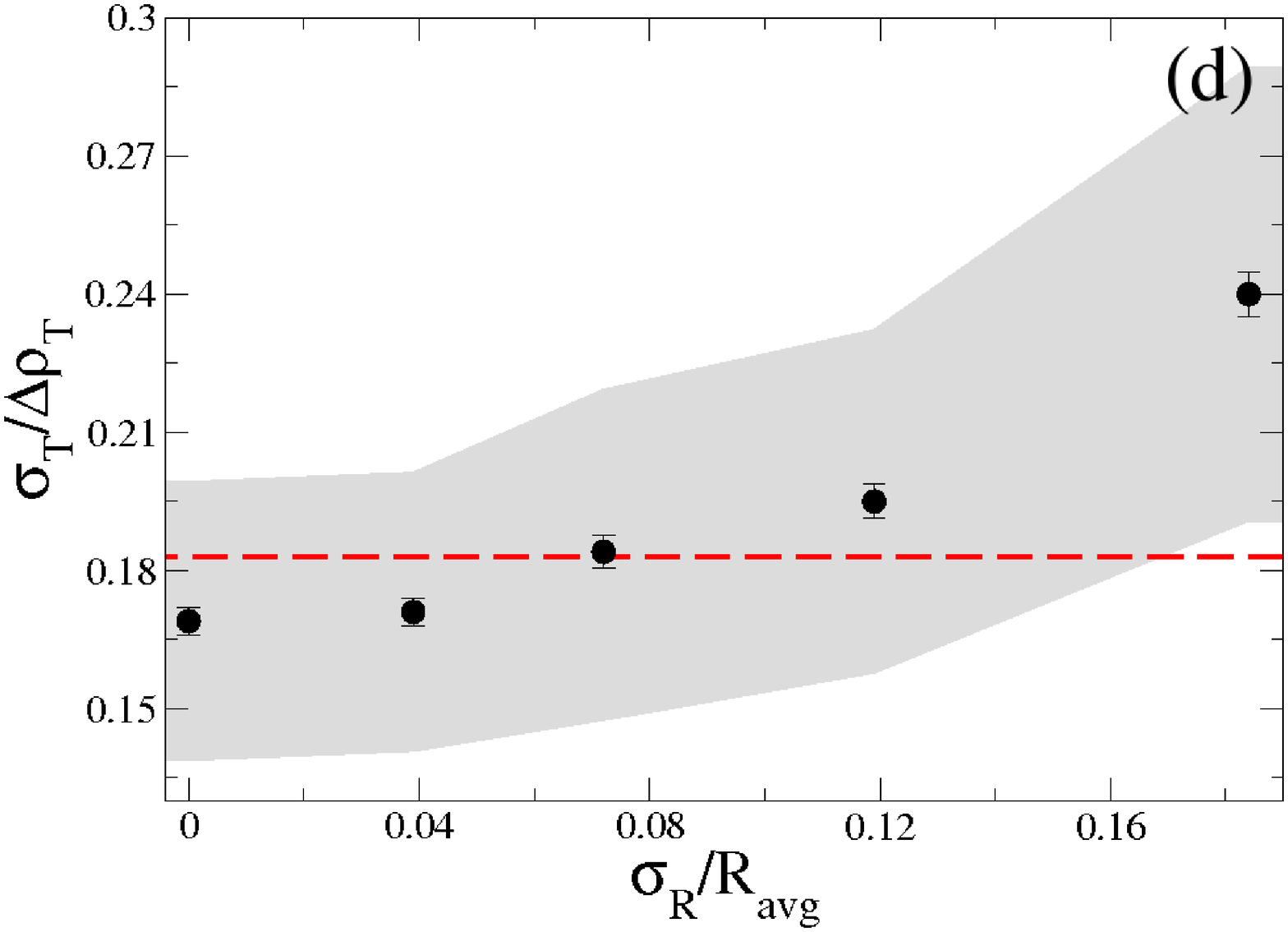}
  \end{tabular} 
  \end{center}
  \caption{\label{fig:popa}(a)  non-stochastic concentrations after $t=4$h of labeled fixed-nitrogen (fN) exposure vs. the cell index $i$. Heterocysts are indicated by  (``H''). Shown are the total fN ($\rho_T$, blue solid line with circles using the left axis) and the free fN ($\rho_F$, red solid line with squares using the right axis).  Note the dramatic difference in scale. The steady-state analytic result for the free fN from a continuum approximation, Eqn.~\ref{eq:free}, is shown as a green line; while the corresponding plateau in $\rho_T$ due to incorporated fN is shown by the black dashed line.  (b)  averaged stochastic $\rho_T$ using doubling-time variation $\Delta$ = 4.5h, with error bars too small to be seen outside data points. The average is over 1000 distributions.  (c) the experimental fN concentrations of Popa \emph{et al} \cite{popa07} in red circles (scale on left axis) and from a single stochastic result from our model using $\Delta$ = 4.5h and $G_{het}$ = 4.25$\times$10$^6$ s$^{-1}$ with $\rho_T$ in blue squares and 25$\rho_F$ in green diamonds (scale on right axis, offset for clarity). Also shown as black lines are the vegetative cell averages. (d) the coefficient of variation in  vegetative cell fN levels, $\sigma_T/\Delta\rho_T$, vs. the coefficient of variation in the growth rate of the vegetative cells, $\sigma_R/R_{avg}$.  The error bars indicate statistical errors of the average variation measured over 100 samples. The grey shaded region indicates the size of the standard deviation, which corresponds to the expected error for a single sample. The red dashed line indicates the variation in the experimental data of Popa \emph{et al} \cite{popa07}, with a single sample. For (a)-(d), we have used only cytoplasmic transport with $D_C$ = 1.54 $\mu$m s$^{-1}$ and $D_I$ = $D_E$ = $D_P$ = 0.}
\end{figure*}

As seen in the last section, for $t \gtrsim 120s$ the free fN distribution has reached an approximate steady-state. We can obtain an analytic expression for this steady-state distribution of $\rho_F$ by approximating the filament as a continuous diffusive medium. The diffusion and consumption of free fN within a continuous filament is  
\begin{equation}
	\partial \rho_F/ \partial t=D \nabla^2\rho_F-c
	\label{eq:diff}
\end{equation}
with $\rho_F$ as the free fN concentration; $D$ is the diffusivity and $c$ is the fN consumed by growth per unit length. For the steady state ($\partial \rho_F/\partial t=0$) the general solution is quadratic (parabolic) vs. distance $x$ from a nearby heterocyst.  In steady-state, the flux on either side of  the heterocyst at $x=0$ equals half of the fN production, i.e. $G_{het}/2 = -D \partial \rho_F/\partial x$.  This flux, $G_{het}/2$, must also equal the total fN consumed by the growing vegetative cells on either side of it, i.e. $c x_0$ where $x_0$ is where growth halts and $\rho_F(x_0)=0$.  Applying $\rho_F(G_{het}/(2 c))=0$, and $-D \partial \rho_F/\partial x (0)=G_{het}/2$ gives us the steady-state profile
\begin{equation}
	\label{eq:free}
	\rho_F(x) =\frac{1}{2}\frac{c}{D}x^2-\frac{G_{het}}{2D}x+\frac{G_{het}^2}{8cD},
\end{equation}
for $x \leq x_0$ and $\rho_F(x)=0$ for $x \geq x_0$. This is in excellent agreement with the steady-state $\rho_F$ from our numerical model with non-stochastic cytoplasmic-only transport, as shown in Figure \ref{fig:popa}(a). 

In our non-stochastic model all cells with free fN levels above zero grow at the same rate and so incorporate fN at the same rate.  This produces a constant plateau in $\rho_I$ for vegetative cells between the heterocyst at $x=0$ and the cell where the free fN vanishes, i.e. for $0<x<x_0$. This is  seen in our discrete cellular model, as shown with $\rho_T$ in Fig.~\ref{fig:popa}(a). We note that the cells at $i=9$ and $i=11$  grow at a reduced rate despite having $\rho_F=0$, since some free fN is provided from their neighbours --- i.e. $\Phi_{in}>0$ in Eq.~\ref{eq:requation}. The plateau height depends upon how much time has elapsed since labeled fN was introduced. In Fig.~\ref{fig:popa}(a) we show results for $t=4$h, to be able to directly compare with the data of Popa \cite{popa07}. The predicted plateau height ($<$$R^{opt}$$>$$gt/$$<$$L$$>$) from our previous continuum model is shown with a black dashed line, and the agreement is excellent. We also see that the free fixed-nitrogen $\rho_F$ (scale on right side) is approximately two orders of magnitude smaller than the total fixed-nitrogen $\rho_T$ (scale on left side), which is dominated by the incorporated fixed-nitrogen $\rho_I$.  We see excellent qualitative agreement with the plateau-like distribution seen by Popa. The expected gradient of free fixed-nitrogen is too small to be easily resolved by their technique, compared to the much larger plateau of incorporated fixed-nitrogen.

\subsection{Stochasticity of Long Time Distributions}
\label{subsec:longtimestochastic}

Fig.~\ref{fig:popa}(b) shows the total fN distribution using our stochastic model that has been averaged over 1000 vegetative segments between two heterocysts. Qualitatively it looks like our deterministic model results. In (c) we show both the experimental results of Popa \emph{et al} \cite{popa07} (red curve, left axis) and one single sample distribution from our stochastic simulation (blue curve, right axis, with an arbitrary factor to separate the curves). Qualitatively the distributions are very similar, but strikingly rougher than the smooth curves in (a) and (b).  The $\rho_F$ in (c) is smooth and relatively small in magnitude, indicating that the variation is almost entirely due to incorporated fN.

In our stochastic simulation, the variation of $\rho_T$ shown in (c) is due both to the varying growth rates of cells along the filament and to their varying lengths --- which leads to corresponding differences in the density of incorporated fN after a fixed time. This raises the possibility that the variability seen by Popa \emph{et al} \cite{popa07} is not due to instrumental or sampling noise, but simply reflects growth rate variability in the filament. In (d) we plot the coefficient of variation of the total-fixed nitrogen, $\sigma_T/\Delta \rho_T$, vs the coefficient of growth rate variation, $\sigma_R/R_{avg}$. Here $\sigma_R$ is the standard deviation of the growth rate distribution, and $R_{avg}$ is the average growth rate. Similarly $\sigma_T$ is the standard deviation of $\rho_T$ in vegetative cells, while $\Delta \rho_T$ is the difference between the average $\rho_T$ in vegetative cells and in heterocysts. While $\sigma_R/R_{avg}$ was calculated exactly, we measured $\sigma_T/\Delta \rho_T$ over 100 segments of cells, and used this to determine the average coefficient of variance, the statistical errors, and the standard deviation around the average --- as shown in (d).  With the dashed red line, we also show $\sigma_T/\Delta \rho_T$ from the Popa experimental data in (c). The grey region shows one standard deviation, which approximates the single interval error for the Popa data. We see that a broad range of growth rate stochasticities are consistent with the Popa's observed variation, including the coefficient of variation $\sigma_R/R_{avg} = 0.165$ reported for a mutant of PCC7120  \cite{allard07}.  We conclude that the variability seen by Popa \emph{et al} is probably determined by the stochasticity of the growth rate. 

After extended isotopic dinitrogen exposure, much longer than typical cell-doubling times, we expect a nearly constant distribution of $\rho_T$ --- corresponding to the nearly uniform concentration of fN in a cyanobacterial filament.  Notably, the variation within the plateau, evident in the data of Popa \emph{et al} \cite{popa07} and modelled in Fig.~\ref{fig:popa}(d), will also vanish at late times --- since it arose due to differential growth rates leading to variations in labeled vs. unlabelled $\rho_I$. Only old heterocysts, which retain some unlabelled fN, will present local dips in the concentration profile. On top of this nearly constant $\rho_I$, will be much smaller steady-state free fN gradients coming from heterocysts --- as described by Eq.~\ref{eq:free} and illustrated in Fig.~\ref{fig:popa}(a-c).   Imaging fN distributions in filaments that have been grown for weeks in labeled dinitrogen may allow small gradients of $\rho_F$ to be discerned with NanoSIMS techniques \cite{popa07}.

\subsection{Long Time Nitrogen Distributions With Periplasmic Transport}
\label{subsec:periplasmic}

We have established that the qualitative characteristics of both the spiked of Wolk \emph{et al} \cite{wolk74}  and the flat distribution of Popa \emph{et al} \cite{popa07} can be reproduced using a single quantitative model for the cyanobacterial filament using only cytoplasmic transport. In this section, we  explore the question of how much additional periplasmic transport is possible while retaining the qualitative phenomenology.  The strongest effect is seen after a Popa-like plateau is formed in $\rho_T$.

\begin{figure*}
  \begin{center}
  \begin{tabular}{cc}
  \includegraphics[width=3.0in]{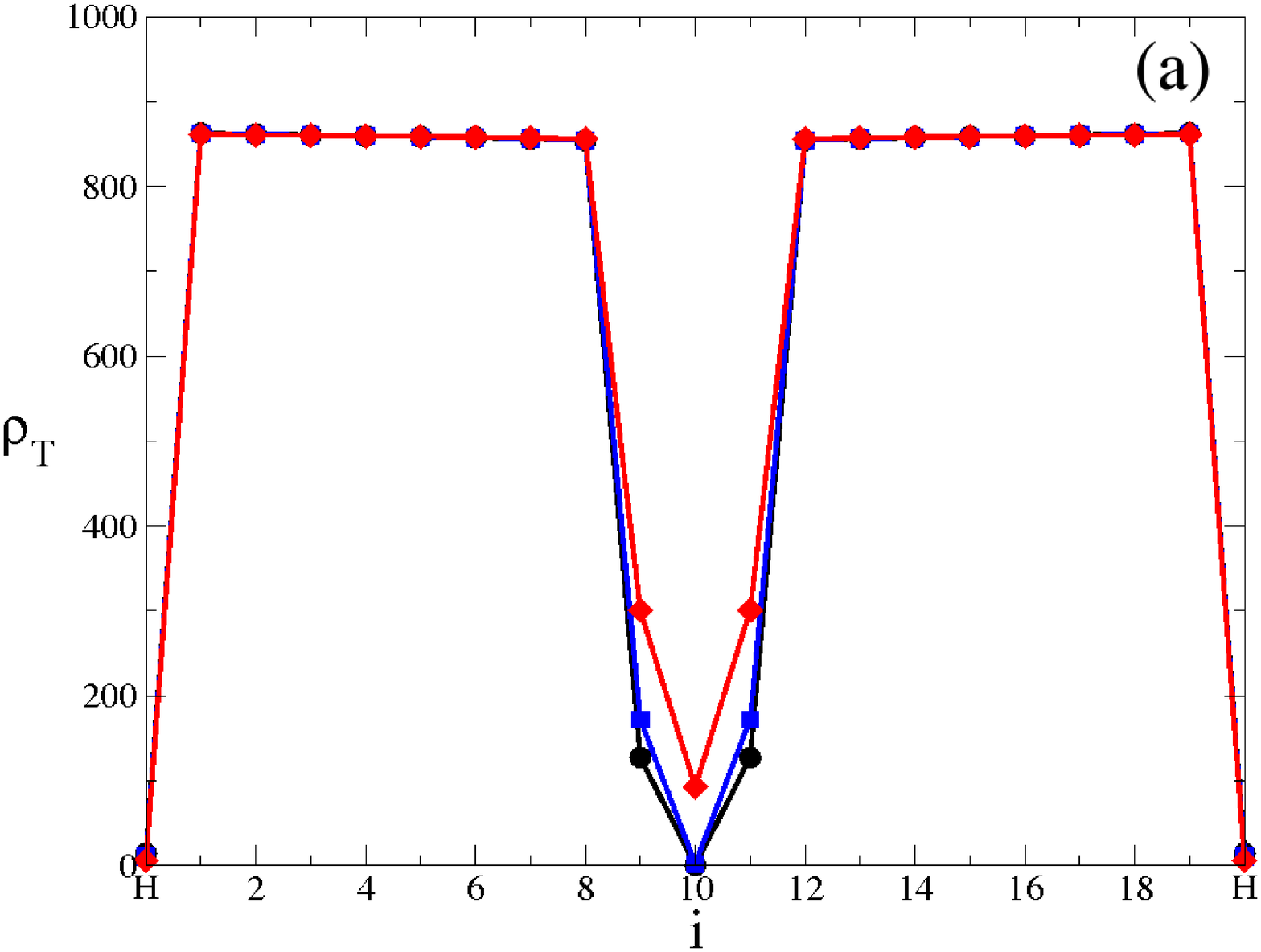}&
  \includegraphics[width=3.0in]{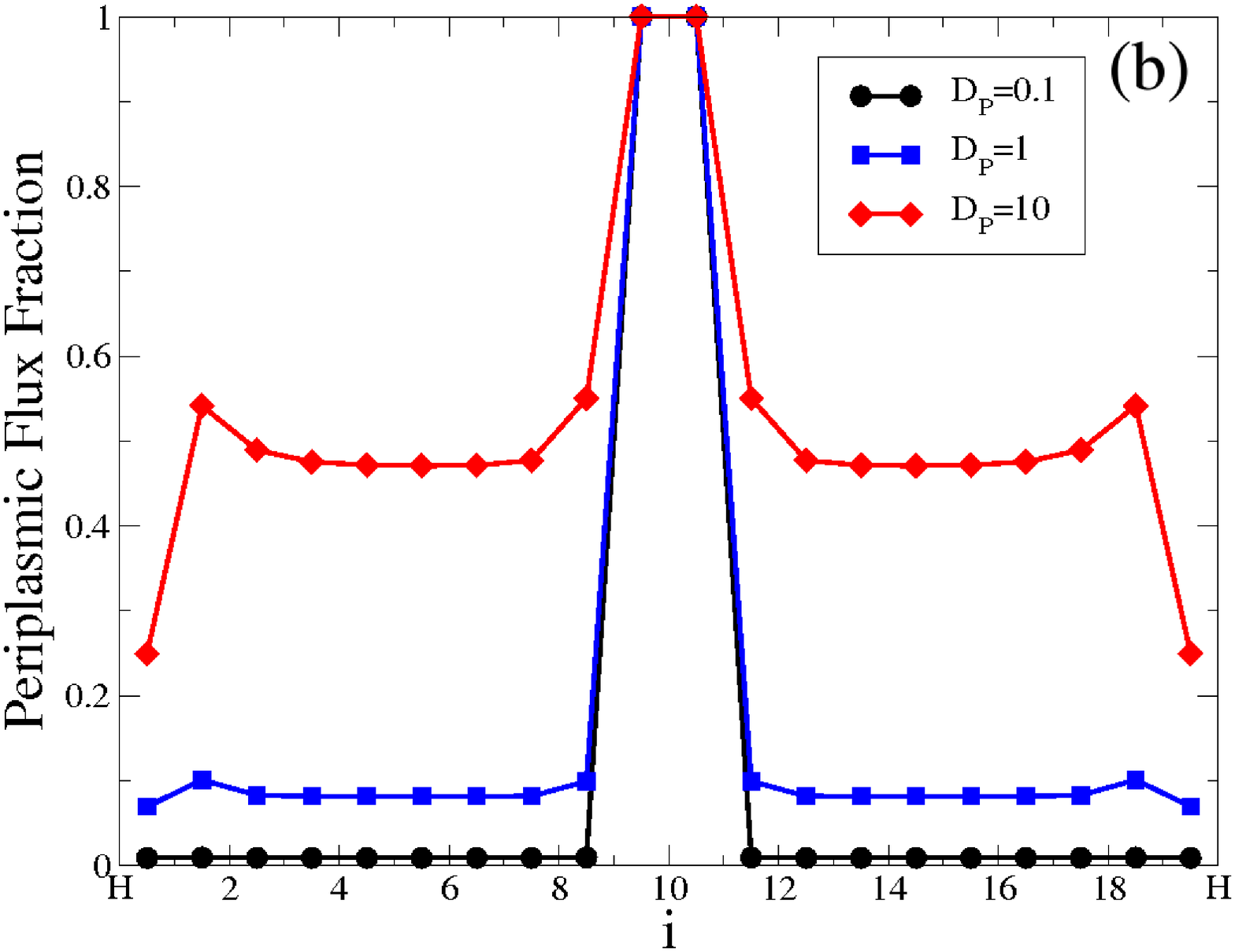}\\
    \includegraphics[width=3.0in]{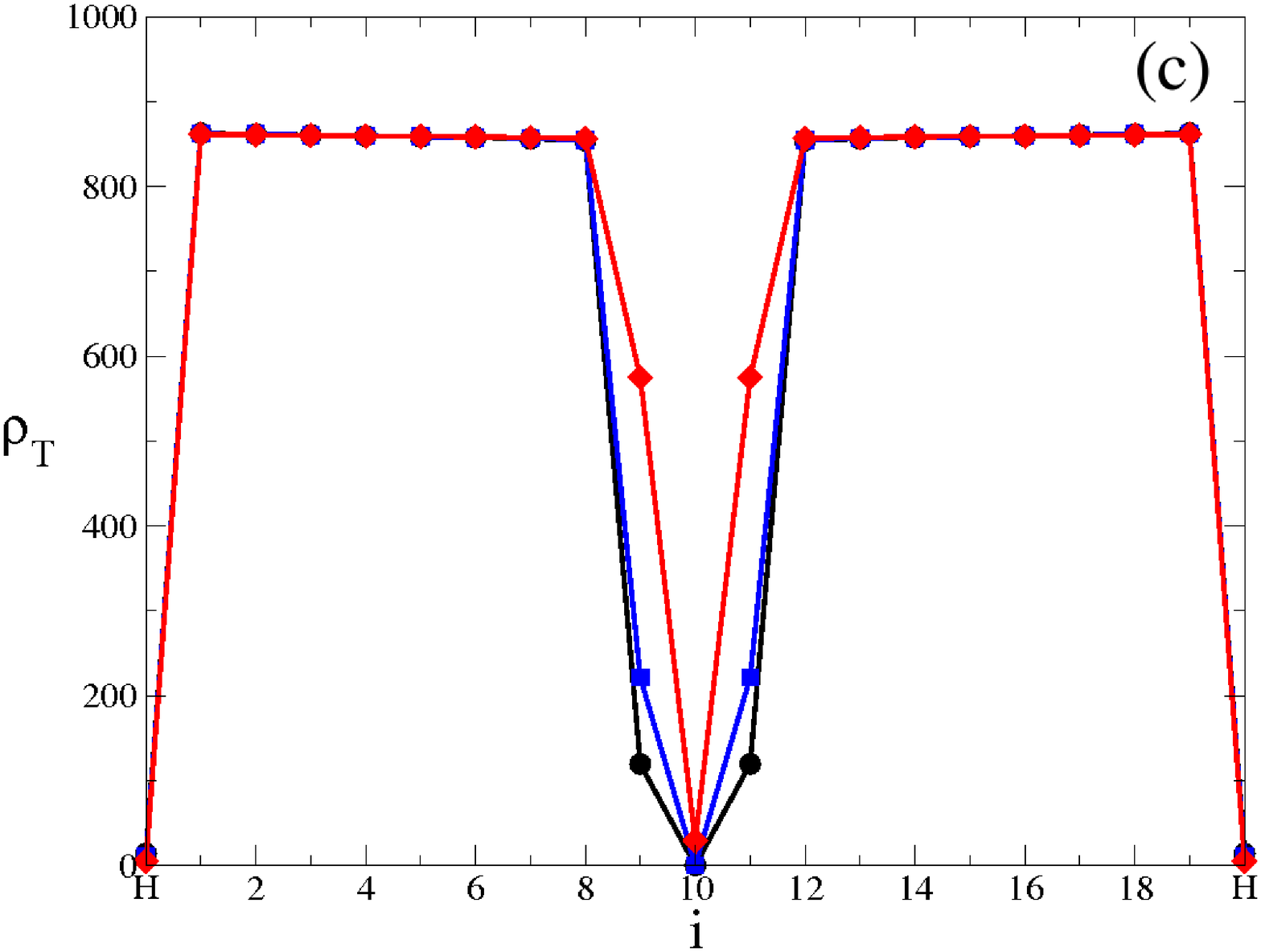}&
    \includegraphics[width=3.0in]{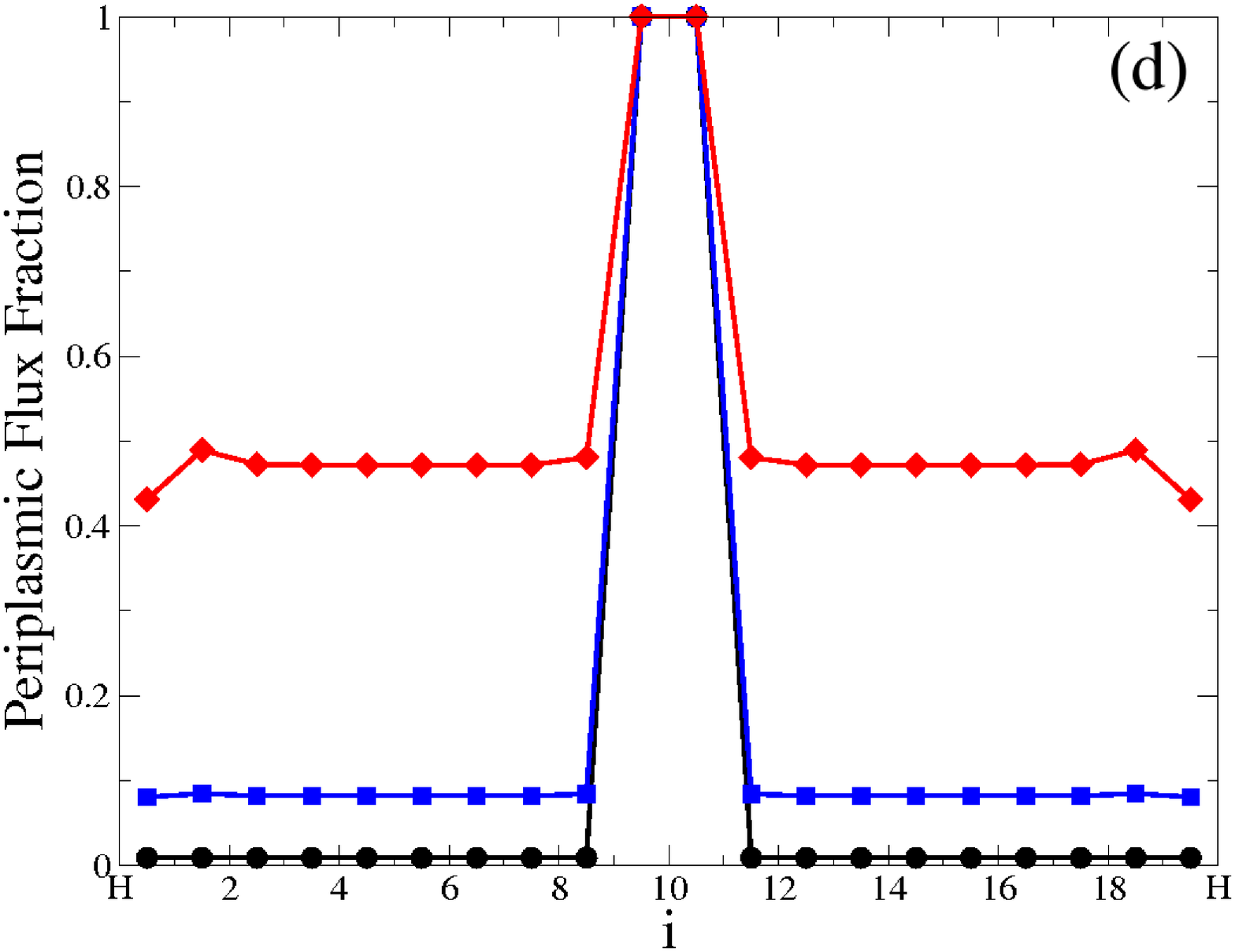}\\
  \end{tabular}
  \end{center}
  \caption{\label{fig:fluxes} Fixed-nitrogen distributions for systems with additional periplasmic transport, as discussed in Section~\ref{subsec:periplasmic}. In all four plots the periplasmic transport rate $D_P$ (in units of $\mu$m s$^{-1}$) is varied according to the legend in (b).  (a)  the total fixed-nitrogen distribution, $\rho_T$, for transport between the cytoplasm and periplasm with $D_I$ = 4.98 s$^{-1}$ and $D_E$ = 0.498 s$^{-1}$. The data of (b) corresponds to the curves of the same colour in (a) and shows the fraction of the total fixed-nitrogen flux along the filament that is periplasmic vs. the cell index $i$. (c) and (d) are similar to (a) and (b), respectively, except that import and export between the cytoplasm and periplasm are tenfold stronger, with  $D_I$ = 49.8 s$^{-1}$ and $D_E$ = 4.98 s$^{-1}$. As before, $D_C$=1.54 $\mu$m s$^{-1}$.  We use $D_L= 0.0498$s$^{-1}$ and $G_{het}$ = 3.67$\times$10$^6$ s$^{-1}$.}
\end{figure*}

In Figs.~\ref{fig:fluxes}(a) and (b) import and export between the cytoplasm and periplasm has $D_I$ = 4.98 s$^{-1}$ and $D_E$ = 0.498 s$^{-1}$ (see Appendix B). The only parameter that varies is the value of the periplasmic diffusivity $D_P$, as indicated by the legend in (b). We vary $D_P$ up to the magnitude expected for unimpeded diffusive transport of glutamine along the filament, as discussed in Sec.~\ref{subsec:dynamics}. The characteristic dip in the centre of the total fixed-nitrogen distribution, $\rho_T$, in (a) fills in at larger values of $D_P$ --- but maintains its qualitative features throughout the range. The mid-segment dip fills in because periplasmic transport allows free fixed-nitrogen to move past the first starving vegetative cell with $N_C=0$. To evaluate how significant the periplasmic transport is, we show in (b) the fraction of the flux along the periplasm to the total flux along the cytoplasm and periplasm. The flux is mostly cytoplasmic next to the heterocysts, and mostly periplasmic next to starving cells.  In between, up to 50$\%$ of the flux can be periplasmic. For the larger $D_I$ and $D_E$ values of Figs.~\ref{fig:fluxes}(c) and (d) the results are very similar. We conclude that significant, though probably not dominant, periplasmic transport is consistent with the qualitative fixed-nitrogen distribution reported by Popa \emph{et al} \cite{popa07}.

\section{Summary}
\label{sec:summary}

We have presented a quantitative model for fixed-nitrogen (fN) transport and dynamics in a cyanobacterial filament, including growth of vegetative cells and production of fN by heterocysts. Our model reproduces the qualitative fN distribution near heterocysts (Fig.~\ref{fig:wolk}) seen after short exposure to isotopically labeled dinitrogen by Wolk \emph{et al.} \cite{wolk74} (``Wolk''); but also reproduces the qualitative distribution between heterocysts (Fig.~\ref{fig:popa}) seen with nanoSIMS techniques after much longer exposure by Popa \emph{et al.} \cite{popa07} (``Popa'').  The apparent lack of a fN gradient in the Popa distribution is explained by the much larger amount of incorporated fixed-nitrogen, which gives the characteristic noisy plateau between heterocysts. Although much smaller in magnitude, the modelled free fN distribution does show a smooth gradient away from heterocysts (Fig.~\ref{fig:popa}(a)) reflecting diffusive transport. 

Qualitatively, our results indicate the importance of the large, noisy, plateau-like concentration profile of incorporated fixed-nitrogen. The level of the plateau  will  increase linearly in time  after the free fN gradient away from heterocysts reaches its steady-state, which takes minutes. The free fN, $\rho_F$, is soon much smaller than the incorporated fN, $\rho_I$. The spike in the Wolk distribution, seen in the first few minutes of exposure to isotopically labeled dinitrogen, was when the free fN still dominated the total $\rho_T$. In contrast, the plateau seen in the Popa distribution was after hours of exposure, when the total fN was dominated by the incorporated component and the gradient in $\rho_F$ was relatively small even with respect to variations of $\rho_I$ within the plateau (see Fig.~\ref{fig:popa}(c)).  At these intermediate times, long after the steady-state $\rho_F$ but much shorter than typical division times, the heterocysts present dips of labeled fixed-nitrogen --- as  seen by Popa \cite{popa07}. Stochasticity, through initial cell lengths and cellular growth rates, adds noise to the incorporated fN distribution, $\rho_I$. As we show in Fig.~\ref{fig:popa} (d), the modelled cell-to-cell variation is due to differences in growth rate which manifest themselves as different amounts of labeled fN incorporated by growth. We believe this explains the observed variation in the experimental nitrogen distribution of Popa \emph{et al} \cite{popa07}.

The possibility of periplasmic fixed-nitrogen transport raised by a contiguous periplasm \cite{flores06}, transport of GFP \cite{mariscal07}, and an outer membrane permeability barrier \cite{nicolaisen09}, led us to explore the possible impact of periplasmic transport in addition to cytoplasmic cell-to-cell connections \cite{mullineaux08}. As illustrated in Fig.~\ref{fig:fluxes}, we found that a significant proportion (up to 50$\%$) of fN transport along the filament can be observed without qualitatively changing our fN distributions. We cannot rule out a role for the periplasm in fN transport, and larger proportions may be achievable with more extensive parameter searches. Indeed, the fact that knocking out amino-acid permeases leads to impaired diazotrophic growth \cite{montesinos95,picossi05,pernil08} does indicate that significant leakage from the cytoplasm does occur (i.e. $D_E>0$) --- which is a necessary part of periplasmic transport.  Nevertheless, our results are also consistent with no significant periplasmic transport.  

We have used a simple model, Eq.~\ref{eq:requation}, to limit growth in response to local fixed-nitrogen starvation: cells grow at a fixed rate $R^{opt}$ if there is sufficient free fixed-nitrogen available locally ($N_C>0$), and at the maximal rate allowed by local input fluxes of free fixed-nitrogen if not. This model reflects the absolute limitation on growth placed by available fixed-nitrogen, but also recovers the characteristic plateau of $\rho_T$ seen in the Popa distribution. As we illustrate in \ref{app:growth}, ``graded'' growth rates, that smoothly depend on $N_C$ and vanish at $N_C=0$, respond to the graded distribution of free fixed-nitrogen away from heterocysts, Eq.~\ref{eq:exp}, and produce graded growth patterns and graded distributions of $\rho_I$ and $\rho_T$ (see Fig.~\ref{fig:wolkproportional}) --- unlike those observed by Popa \cite{popa07}. Graded growth models, such as the Monod model \cite{kovarova-kovar98}, are typically designed to describe response to external metabolite concentrations rather than cytoplasmic concentrations. Like our growth model,  \emph{Salmonella typhimurium} \cite{ikeda96} and  \emph{Escherichia coli} \cite{brauer06,hart11} both show non-graded growth in response to cytoplasmic fixed-nitrogen starvation.  However, there is little direct evidence of how growth of vegetative cells in cyanobacterial filaments depends on the cytoplasmic freely available fixed-nitrogen, $\rho_F$. 

The experimental work was done on different species of filamentous cyanobacteria, namely \emph{Anabaena cylindrica} by Wolk \emph{et al} \cite{wolk74} and \emph{Anabaena oscillarioides} by Popa \emph{et al} \cite{popa07}.  In contrast, our models were parameterized by consideration of  transport studies done in \emph{Anabaena} sp. strain PCC7120 (see Sec.~\ref{subsec:dynamics}). While we expect  parameter values to differ  between species, so that precise numerical agreement with the Wolk and Popa results is not to be expected, our qualitative agreement indicates that similar fixed-nitrogen transport and dynamics could apply in these different model organisms.  

We expect the qualitative features of the Wolk and Popa fixed-nitrogen distributions to apply to the early and late-time fixed-nitrogen distributions, respectively, of all filamentous cyanobacteria with heterocysts that are growing in media lacking fixed-nitrogen and that have significant cytoplasmic connections.  Nevertheless, quantitative details will depend, via the model parameterization, on the cyanobacterial strain and the experimental conditions.  For example, sufficiently large periplasmic transport (via $D_P$) would broaden Popa's dip in incorporated fixed-nitrogen between heterocysts due to transport and growth past the first starving cell.  Conversely, a larger cytoplasmic diffusivity (via $D_C$) would alter the magnitude of Wolk's free fixed-nitrogen (via Eqn.~\ref{eq:free}) but not the incorporated fixed-nitrogen.  The average doubling time determines the characteristic time at which we expect the later time distribution to emerge. Cell-to-cell variations in the doubling time, in turn, determine the amount of variation expected within the Popa-like plateau of incorporated fixed-nitrogen near heterocysts for the first few generations of labelled growth.

Our results reconcile the metabolite patterns shown by the autoradiographic technique of Wolk \emph{et al} \cite{wolk74} with the nanoSIMS technique of Popa \emph{et al} \cite{popa07}. It is not obvious which would currently have the best resolution, and it would be useful to have either or both applied to \emph{Anabaena} PCC7120 --- the most popular current model for filamentous cyanobacteria and for which recent transport studies exist \cite{mullineaux08,mariscal07}.  We can make four qualitative but testable predictions from our work.  First, as the duration of labelled fN is increased the total fN distribution away from heterocysts will evolve from a Wolk-like peak, dominated by free fN, to a Popa-like plateau, dominated by incorporated fN.  As indicated by Fig.~\ref{fig:wolk}, the crossover takes place after only a few minutes of labeled dinitrogen fixation. Second, on a similar timescale the free fN will  reach a steady-state smooth parabolic shape even in a single sample. This free fN may be directly measurable, for example, by fluorescence resonance energy transfer (FRET)-based metabolite nanosensors \cite{fehr04}. The parabolic gradient of $\rho_F$ could also be uncovered by averaging the results of many experiments measuring $\rho_T$. Third, the variations seen by Popa \emph{et al} \cite{popa07} will decrease in relative magnitude after at least one doubling-time of exposure to labeled fN.    Fourth is more of an assumption, but we expect that growth of vegetative cells will slow abruptly when the freely available fN vanishes --- according to Eqn.~\ref{eq:requation} rather than in a graded fashion according to Eqn.~\ref{eq:graded}.  Finally, with more precise parameterization of our transport parameters, sources, and sinks we can make quantitatively testable predictions. Prediction of the variance of cell-to-cell fN, $\sigma_I$, requires only the observed growth rate variance $\sigma_R$ and the duration of exposure to labeled dinitrogen.

\ack
We thank the Natural Science and Engineering Research Council (NSERC) for support, and the Atlantic Computational Excellence Network (ACEnet) for computational resources. AIB also thanks NSERC, ACEnet, and the Killam Trusts for fellowship support.

\pagebreak

\appendix
\section{Graded growth models}
\label{app:growth}

In this Appendix we address the effects of a graded growth model, where the growth rate of a cell depends continuously on the cytoplasmic free fixed-nitrogen concentration $\rho_F= N_C/L$, and vanishes when $N_C=0$. Examples of such graded growth models include the Monod model (normally applied to concentrations of extracellular metabolites) \cite{kovarova-kovar98} and the growth model used by Wolk \emph{et al} \cite{wolk74} for modelling the response of cyanobacterial filaments to short exposures to labeled dinitrogen.  Here we consider the model of Wolk \emph{et al} \cite{wolk74}, though we expect similar results with any graded growth model.

The graded-growth model of Wolk \emph{et al} \cite{wolk74} has growth linearly proportional to free fixed-nitrogen concentration, $\rho_F$. The dynamics of $\rho_F$ are then
\begin{equation}
\label{eq:gradedde}
	\partial \rho_F/\partial t=D\nabla^2\rho_F-k\rho_F,
\end{equation}
where the first term on the right represents diffusion along the filament and the second term is the local sink due to vegetative growth.  (Note the difference with our coarse-grained growth model in Eq.~\ref{eq:diff}.) We ignore the discrete cellular structure of the filament in order to obtain an analytic solution. A heterocyst at $x=0$ will impose a flux $J= -D \partial \rho_F/\partial x = G_{het}/2$ there, leading to a steady-state solution 
\begin{equation}
  	\rho_F=\frac{1}{2\sqrt{kD}}e^{-\sqrt{k/D}x}.
	\label{eq:exp}
\end{equation}
The incorporated fixed-nitrogen is proportional to this, and so $\rho_T$ will exponentially decay away from the heterocyst with a length-scale $\sqrt{D/k}$. This is not observed in the distribution of Popa \emph{et al} \cite{popa07} --- since the plateau shown in Fig.~\ref{fig:popa} is inconsistent with a small $\sqrt{D/k}$ while the sharp dip seen between existing heterocysts is inconsistent with a large $\sqrt{D/k}$.  

\begin{figure}[h]
  \begin{center}
  \begin{tabular}{c}
    \includegraphics[width=3.35in]{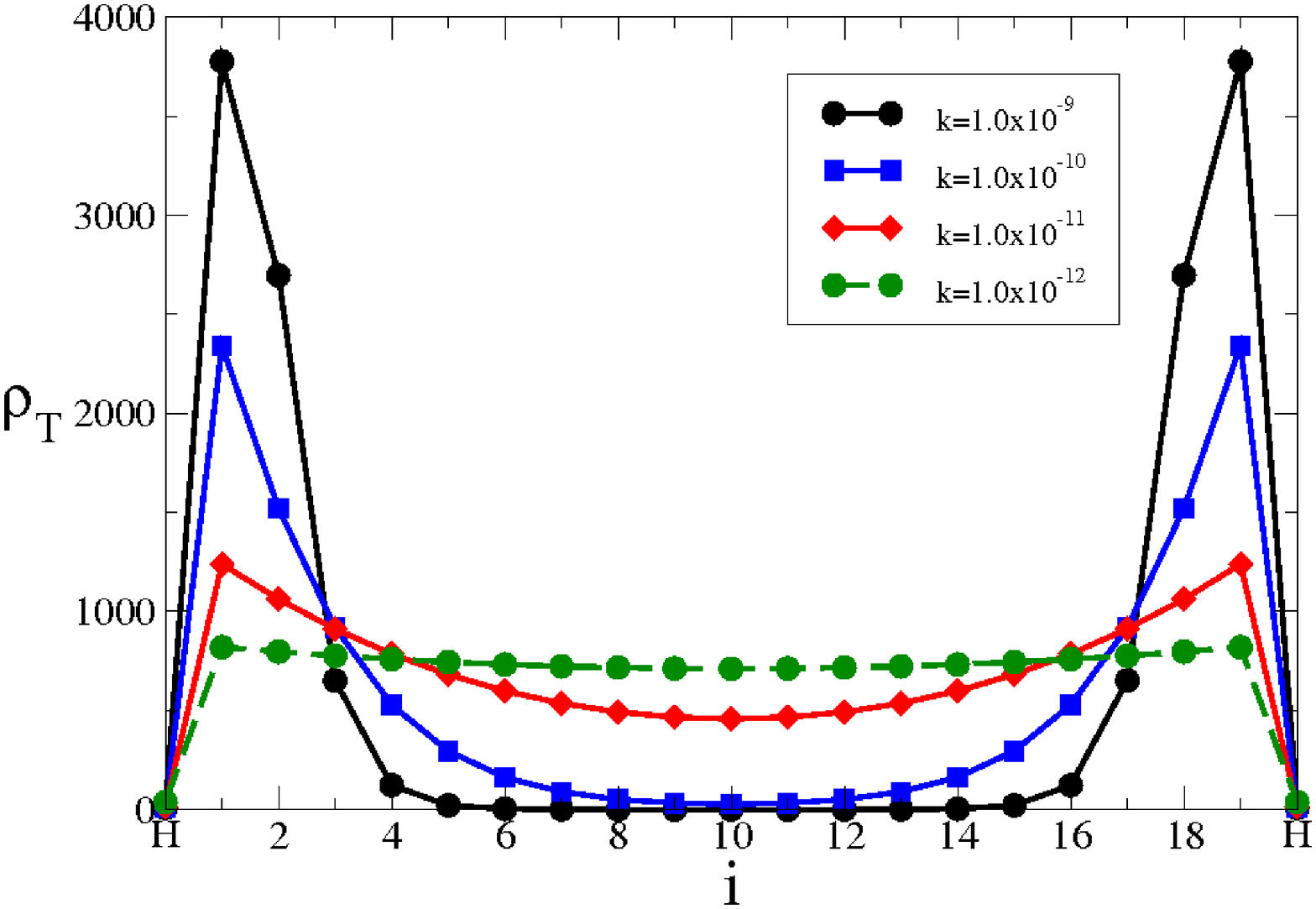}\\
  \end{tabular}
  \end{center}
  \caption{\label{fig:wolkproportional} The numerical total fixed-nitrogen distributions vs. cell index $i$ using graded growth proportional to local $\rho_F$ as Eq.~\ref{eq:graded}. The growth-rate proportionality constant $k$ (in units of $\mu$m s$^{-1}$) is varied, as indicated by the legend. This is a deterministic cellular model with cytoplasmic transport: $D_C$ = 1.54 $\mu$m s$^{-1}$ and $D_I$ = $D_E$ = $D_P$ = 0.}
\end{figure} 

With our discrete deterministic model, we have modelled the Wolk-style growth 
\begin{equation}
R_{graded}(N_c) = k N_c/L_i.
\label{eq:graded}
\end{equation}
The results are shown in Fig.~\ref{fig:wolkproportional} for various values of $k$. The discrete nature of the model modifies the observed pattern near the bounding heterocysts compared to Eq.~\ref{eq:exp} --- but either a plateau but no mid-segment dip is recovered (for small $k$) or a wide mid-segment trough with no plateau is recovered (for large $k$). We are unable to recover the experimental distribution seen by Popa \emph{et al} \cite{popa07} with such a graded growth model.

\section{Parameter estimation of $D_I$ and $D_E$}

To estimate the value of $D_I$ (from Eqns. \ref{eq:cytonitrogen} and \ref{eq:perinitrogen}) we begin with the measurement of Pernil \emph{et al} \cite{pernil08} that 1.65 nmol (mg protein)$^{-1}$ is imported into a PCC7120 filament in 10 minutes from a medium containing 1 $\mu$M glutamine. From our discussion in Section \ref{sec:growth}, the dry mass of a PCC7120 cell is 4.89$\times$10$^{-12}$g and the mass of protein, approximated at 55$\%$ of the dry mass of the cell \cite{philips09}, is 2.69$\times$10$^{-12}$g. This implies that there is 4.44$\times$10$^{-9}$ nmol of glutamine, or $N_G$ = 2.67$\times$10$^6$ glutamine molecules, imported in 10 minutes into each cell. The term in Eqn. \ref{eq:cytonitrogen} describing import from the periplasm to the cytoplasm is $\partial N_C/\partial t$=$D_IN_P$. If we assume that the periplasmic concentration is equal to the glutamine concentration in the external medium, then $N_P$ can be replaced by the product of the glutamine concentration in the periplasm, $\rho_P$, the cross sectional area of the periplasm $A$ (for a 0.1$\mu$m thick periplasm surrounding a cytoplasm 1$\mu$m in radius $A=0.21\pi\mu$m$^2$), and the cell length $L$ giving $N_G=D_I\rho_PLA\tau$, with $\tau= 10$ minutes. Solving this for $D_I$ using $L$ = $L_{min}$ = 2.25$\mu$m yields $D_I$ = 4.98 s$^{-1}$. ABC transporters, known to transport amino acids \cite{picossi05}, are asymmetric, and we assume that export is 10$\times$ weaker than import, and that $D_E$ = 0.498 s$^{-1}$. It is also known the the outer cell membrane provides a permeability barrier \cite{nicolaisen09}, implying that the periplasmic glutamine concentration may not be as high as the external glutamine concentration, and so we  also consider a $D_E$ and $D_I$ pair that is ten times larger with $D_I$ = 49.8 s$^{-1}$ and $D_E$ = 4.98 s$^{-1}$.

\pagebreak

\section*{References}
\bibliographystyle{unsrt}
\bibliography{essay2}

\end{document}